\documentclass{ckm}                 

\usepackage{txfonts}            

\confname{Workshop on the CKM Unitarity Triangle, IPPP Durham, April
  2003}

\newcommand{\beq}{\begin{equation}}
\newcommand{\eeq}{\end{equation}}
\newcommand{\beqn}{\begin{eqnarray}}
\newcommand{\eeqn}{\end{eqnarray}}
\newcommand{\vs}{\\[0.3cm]\noindent}
\def\EPJ{{\em Eur. Phys. J.}}
\def\NP{{\em Nucl. Phys.}}
\def\PRL{{\em Phys. Rev. Lett.}}
\def\PRD{{\em Phys. Rev.}}
\def\ea{{\em et al.}}
\def\PKzpi{P_{K\pi}^{0+}}
\def\Tpipi{T^{+-}}
\def\Ppipi{P^{+-}}
\def\deltapipi{\delta_{\pi\pi}}

\def\CP{{\em CP}}
\def\CPmath{CP}
\def\bbar{B^0 {\overline B}^0}
\def\Cpipi{C_{\pi\pi}}
\def\Spipi{S_{\pi\pi}}
\def\rar{\rightarrow}
\def\rhobar{\bar\rho}
\def\etabar{\bar\eta}
\def\sta{{\rm sin}\,2\alpha}
\def\stb{{\rm sin}\,2\beta}
\def\a{\alpha}
\def\alphaeff{\alpha_{\rm eff}}
\def\dalpha{\alpha-\alpha_{\rm eff}}

\def\dmt{\Delta t}
\def\dmd{\Delta m_d}
\def\Rth{R_{\rm th}}
\def\babar{\mbox{\sl B\hspace{-0.4em} {\small\sl A}\hspace{-0.37em} \sl B\hspace{-0.4em} {\small\sl A\hspace{-0.02em}R}}}

\def\BRKpi{{\cal B}_{K\pi}^{+-}}
\def\BRpipi{{\cal B}^{+-}}
\def\BRpippiz{{\cal B}^{+0}}
\def\BRpizpiz{{\cal B}^{00}}

\def\Apipi{A^{+-}}
\def\Abarpipi{\overline\Apipi}

\def\rfit{{\em R}fit}
\def\ckmfitter{{\em CKMfitter}}

\def\ie{{\em i.e.}} 

\title{Interpreting \CP\   Asymmetries in \boldmath$B^0\rar \pi^+ \pi^-$ Decays.}
\author{A.~H\"ocker\addressmark{a}, H.~Lacker\addressmark{a}, 
M.~Pivk\addressmark{b} and \underline{L.~Roos}\addressmark{b}}

\address[a]{Laboratoire de l'Acc\'el\'erateur Lin\'eaire,\\
IN2P3-CNRS et Universit\'e de Paris-Sud, BP 34, F-91898 Orsay Cedex, France}
\address[b]{LPNHE-Paris,
IN2P3-CNRS et Universit\'e de Paris VI-VII, Tour 33 rdc, 4 place Jussieu, 
75232 Paris cedex 05, France}

\begin{document}

\begin{abstract}We discuss methods used to interpret the measurements of $CP$-asymmetry 
in $B^0\rar\pi^+\pi^-$ within the framework of the Standard Model. 
 Interpretations using various theoretical inputs are 
given, ranging from the rather general, yet unpredictive, properties
based on strong isospin symmetry to  highly predictive 
calculations using QCD Factorization.  
The calculations are performed using the software package \ckmfitter\
and a frequentist method, \rfit.
\end{abstract}

\maketitle


\section{Introduction}
\label{sec:introduction}

Unlike the measurement of $\stb$ from $B^0\rar J/\psi K^0_S$ or similar
channels, multiple weak phases 
have to be considered in the analyses of charmless $B$ decays aiming at 
the extraction of $\sta$. We discuss here the physical implications 
of the \babar\  and Belle  results on time-dependent \CP-violating 
asymmetries in $B^0\rar\pi^+\pi^-$, taking into account averaged results 
on branching fractions  of $B\rar hh^\prime$ 
decays ($h$, $h^\prime$ denoting neutral or charged pions or kaons). 
The data are interpreted through various frameworks \cite{pipinote}, 
starting from 
theoretical assumptions such as SU(2) invariance and reaching 
highly predictive QCD factorization calculations.

The computational work has been carried out using
the program package \ckmfitter~\cite{CKMfitter}. 
The 
statistical approach \rfit, which is based on a frequentist understanding 
of systematic theoretical uncertainties is used. Theoretical 
parameters are treated as freely varying, bound or unbound,
in the fit.

\section{Theoretical frameworks}

\subsection{Basic Formulae and Definitions}

The complex Standard Model 
amplitudes for $B^0\rar\pi^+\pi^-$ decays have contributions from tree ($T$) 
and penguin ($P$) 
amplitudes with different weak and strong phases, in general. 
Exchange diagrams are included in the tree since
they have the same weak phase. The transition amplitudes
are given by\footnote
{
  We have used unitarity to reorganize the $c$, $u$ and $t$ penguin
  amplitudes according to their weak phases in the tree and penguin
  components of Eqs.~(\ref{eq:b0pi+pi-},\ref{eq:b0barpi+pi-}).
  }
\beqn
\label{eq:b0pi+pi-}
\Apipi(B^0\rar\pi^+\pi^-) 
                        =   |V_{ud}V_{ub}^*| e^{i\gamma}\Tpipi 
                            + |V_{td}V_{tb}^*| e^{-i\beta}\Ppipi \\
\label{eq:b0barpi+pi-}
 \Abarpipi(\overline B^0\rar\pi^+\pi^-) 
                        =   |V_{ud}V_{ub}^*| e^{-i\gamma}\Tpipi 
                            + |V_{td}V_{tb}^*| e^{i\beta}\Ppipi
\eeqn
where $V_{td}=|V_{td}|e^{-i\beta}$ and 
$V_{ub}=|V_{ub}|e^{-i\gamma}$, with $\beta$ and $\gamma$ being the 
angles of the UT. The time-dependent \CP\  asymmetry of the 
physical $\bbar$ system is given by
\beqn
   a_{\CPmath}(\dmt) &\equiv&
        \frac{\Gamma(\overline B^0_{\rm phys}(\dmt)\rar\pi^+\pi^-) 
                   - \Gamma(B^0_{\rm phys}(\dmt)\rar\pi^+\pi^-)}
                  {\Gamma(\overline B^0_{\rm phys}(\dmt)\rar\pi^+\pi^-) 
                   + \Gamma(B^0_{\rm phys}(\dmt)\rar\pi^+\pi^-)}
                \nonumber\\[0.05cm]
        &=& \Spipi {\rm sin}(\dmd \dmt) - \Cpipi {\rm cos}(\dmd \dmt)~,
\eeqn
where $\dmd$ is the $\bbar$ oscillation frequency, $\dmt$ is the 
time difference between the two $B$ decays, and
the coefficients of the sine and cosine terms are given by
\footnote{We assume that \CP\   violation in mixing
is absent, \ie, $|q/p|=1$.}
\beq
\label{eq:spipicpipicoefficients}
   \Spipi = \frac{2{\rm Im}\lambda_{\pi\pi}}{1 + |\lambda_{\pi\pi}|^2}
        ,\hspace{0.05cm}
   \Cpipi = \frac{1 - |\lambda_{\pi\pi}|^2}{1 + |\lambda_{\pi\pi}|^2}
        \hspace{0.05cm} {\rm and} \hspace{0.05cm}
   \lambda_{\pi\pi} 
        = e^{-2i\beta} \frac{\Abarpipi}{\Apipi}
\eeq

Due to the presence of tree and penguin contributions, the phase of $\lambda_{\pi\pi}$ is shifted from $\alpha$
by the relative strong phase $\deltapipi \equiv {\rm arg}(\Ppipi {\Tpipi}^*)$ 
between the penguin and the tree amplitudes. 
An effective angle $\alphaeff$ that incorporates the 
phase shift is defined by $\lambda_{\pi\pi} = 
        |\lambda_{\pi\pi}|e^{2i\alphaeff}$.

In this work, the modulus and the phase of $\frac{\Ppipi}{\Tpipi}$ are 
constrained or calculated within various theoretical frameworks, described 
in the next sections.

\subsection{Constraints from SU(2) symmetry \label{sec:su2}}

Using strong isospin invariance, the amplitudes of the various 
$B\rar\pi\pi$ decays are related to each other. Moreover, 
Gronau and London have shown~\cite{grolon} that the 
measurements of rates and \CP\  asymmetries of the charged and two
neutral $\pi\pi$ final states together with the exploitation of their 
isospin relations provides sufficient information to extract 
$\a$ up to an four-fold ambiguity (within $[0,2\pi]$).

CP-averaged branching fractions ${\cal B}^{ij} = BR(B\rar \pi^i \pi^j)$ 
together with $\Cpipi$ and $\Spipi$ measurements are used here. Using SU(2) 
invariance and assuming that electroweak penguins can be neglected,
one can write  bounds on the relative strong phase $\deltapipi$ and 
$\BRpizpiz$ (\cite{GrQu},~\cite{charles},~\cite{GrLoSiSi}). Among 
them\footnote{The bounds below and in next section are given only to 
understand the results 
presented in section \ref{sec:analysis}, produced with the \ckmfitter fitting 
program.}:

\beq
\label{eq:boundPZ}
   {\rm cos}\,2(\dalpha)\ge\frac{1-2\BRpizpiz/\BRpippiz}{y}
                          + 
                          \frac{\left(\BRpipi-2\BRpippiz+2\BRpizpiz\right)^2}
                               {4\BRpipi\BRpippiz y} 
\eeq
where $y\equiv\sqrt{1-\Cpipi^2}$.

\subsection{Improved bound using SU(3) flavour symmetry \label{sec:su3}}

An additionnal experimental input, the branching fraction of the 
$B^0\rar K^+\pi^-$ decay, is used. 
Under the assumption of SU(3) flavour symmetry and neglecting 
OZI-suppressed penguin annihilation diagrams, which contribute 
to $B^0\rar\pi^+\pi^-$ but not to $B^0\rar K^{+}\pi^{-}$, the 
penguin amplitudes in $B^0\rar\pi^+\pi^-$ and 
$B^0\rar K^{+}\pi^{-}$ are equal, 
which leads to the bound~\cite{charles}:
\beq
\label{eq:Rkpioverpipi}
        {\rm cos}\,2(\dalpha)
        \ge \frac{1-2\lambda^2 \BRKpi/\BRpipi}{y}~
\eeq
where $\lambda$ is the Wolfenstein parameter. 

The flavour symmetry SU(3) is only approximately realized in nature and one 
may expect violations of up to $30\%$ on the amplitudes. 
However, the 
bound in Eq.~(\ref{eq:Rkpioverpipi}) can be considered as conservative, 
 since SU(3) breaking correction could 
strengthen the bound. Indeed, under the assumption of factorization, 
the ratio of branching fractions $\BRKpi/\BRpipi$ 
would be lowered by a factor $(f_{\pi}/f_{K})^{2}\simeq0.67$.

\subsection{Estimating \boldmath$|\Ppipi|$ from $B^{+}\rar K^0\pi^+$ }
\label{sec:naiveFA}

In addition to the isospin relations, the penguin amplitude $|\Ppipi|$ is  
inferred from the branching fraction of 
the penguin-only mode $B^{+}\rar K^0\pi^+$,
$
{\cal B}(B^{+}\rar K^0\pi^{+}) = 
        |V_{tb}^{*}V_{ts}|^2|\PKzpi|^2
$
 by setting~\cite{GrRo} 
$        |\Ppipi| \approx \frac{1}{\sqrt{r_{\tau}}}
                   \frac{f_{\pi}}{f_{K}} \frac{1}{\Rth}
                   |\PKzpi|~.
$

$r_{\tau} = \tau_{B^{+}}/\tau_{B^{0}} = 1.086\pm0.016$~\cite{PDG2002}; $\frac{f_{\pi}}{f_{K}}$
corrects for SU(3) breaking estimated within naive factorization, \ie,
neglecting strong rescattering amplitudes; 
$\Rth=0.98 \pm 0.05$~\cite{BBNS}, introduces a theoretical 
estimate of SU(3) breaking. One has to stress that no theoretical error on 
the various dynamical assumptions, such as the size of non-factorizable 
contributions, or the assumption that $B^+\rar K^0\pi^+$ is given by a pure 
penguin amplitude, is assigned here. 
Note that the strong phase 
$\deltapipi$ remains unconstrained in this method.

\subsection{QCD factorization\label{sec:QCDFA}}

New theoretical methods to calculate the tree and penguin amplitudes in 
$B\rar hh^\prime$ on the basis of QCD have been developed in recent years. 
Such calculations can be used to 
predict the penguin-to-tree amplitude ratio and translate a 
measurement of $\Spipi$ and $\Cpipi$ into a powerful constraint on the CKM 
phases. However, they still need to be validated by 
experimental data. In the present work, the QCD Factorization Approach (QCD FA)~\cite{BBNS}
is  used. If not stated otherwise, the non-factorizable non-calculable 
contributions from annihilation processes and hard spectator interactions 
are fixed to the default values, as defined by the authors (see 
Ref.~\cite{BBNS} for more details).


\section{The Input Data}

\begin{table}[h]
\begin{center}
\setlength{\tabcolsep}{0.8pc}
{\normalsize
\begin{tabular}{lc}\hline
& \\[-0.3cm]
Mode & World average \cite{CPpipiBabar} 
	 	\cite{BRALL}	\\[0.1cm]
\hline
& \\[-0.3cm]
$B^0\rar\pi^+\pi^-$ & $4.78\pm0.54$ 
	 	  
	\\[0.05cm]
$B^+\rar\pi^+\pi^0$ & $5.83\pm0.96$ 
	\\[0.1cm]	
& \\[-0.3cm] 
$B^+\rar\pi^0\pi^0$ & $2.01^{\,+0.70}_{\,-0.67}$ 
	  \\[0.05cm]
& \\[-0.3cm]
$B^0\rar K^+\pi^-$  & $18.46\pm0.98$ 
	\\[0.05cm]
$B^+\rar K^0\pi^+$ & $18.09^{\,+1.73}_{\,-1.69}$ 
	\\[0.05cm]
\hline
\end{tabular}
}
\vspace{0.5cm}
\caption[.]{\label{tab:BRcompilation} \em
	$B\rar hh^\prime$ 
	branching 
        fractions (in units of $10^{-6}$, status as of ICHEP 2002) 
	 
	}
\end{center}
\end{table}

The branching fractions of all $B\rar hh^\prime$ modes used 
in the analysis ($hh^\prime = \pi, K $) 
are given in Table~\ref{tab:BRcompilation}. We also use the results on 
time-dependent \CP\  asymmetry measured by 
\babar~\cite{CPpipiBabar}  and Belle~\cite{CPpipiBelle} 
(statistical and systematic errors have been added in quadrature):
\beqn
\label{eq:resultsCP}
   \begin{array}{lcccc}
	\hline
	&&&&\\[-0.35cm]
		     & \Spipi\ 		& \Cpipi\  	& \rho\\[0.15cm]
	\hline
	&&&&\\[-0.35cm]
	\babar	     &  0.02\pm0.34  	& -0.30\pm0.25 & -0.10 \\[0.05cm]
	 {\rm Belle} & -1.23\pm0.42
					& -0.77\pm0.28
							& 0.024 \\[0.15cm]
	\hline
	&&&&\\[-0.35cm]
	{\rm Average}&  -0.49\pm0.19  	& -0.46\pm0.26  &  0.058 \\
	\chi^2	     &  \multicolumn{2}{c}{7.2\rightarrow2.8\sigma	}
							&  	 \\[0.05cm]

	\hline
   \end{array}
\eeqn
where we have reversed the sign of Belle's $\Cpipi=-A_{\pi\pi}$ to 
account for the different convention adopted. The last column in 
Eq.~(\ref{eq:resultsCP}) gives the statistical correlation coefficients 
between $\Spipi$ and $\Cpipi$ as quoted by the experiments. 

\section{Numerical Analysis of \boldmath$B\rar hh^\prime$ Decays\label{sec:analysis}}

\begin{figure}[h]
\hbox to\hsize{\hss
\includegraphics[width=\hsize]{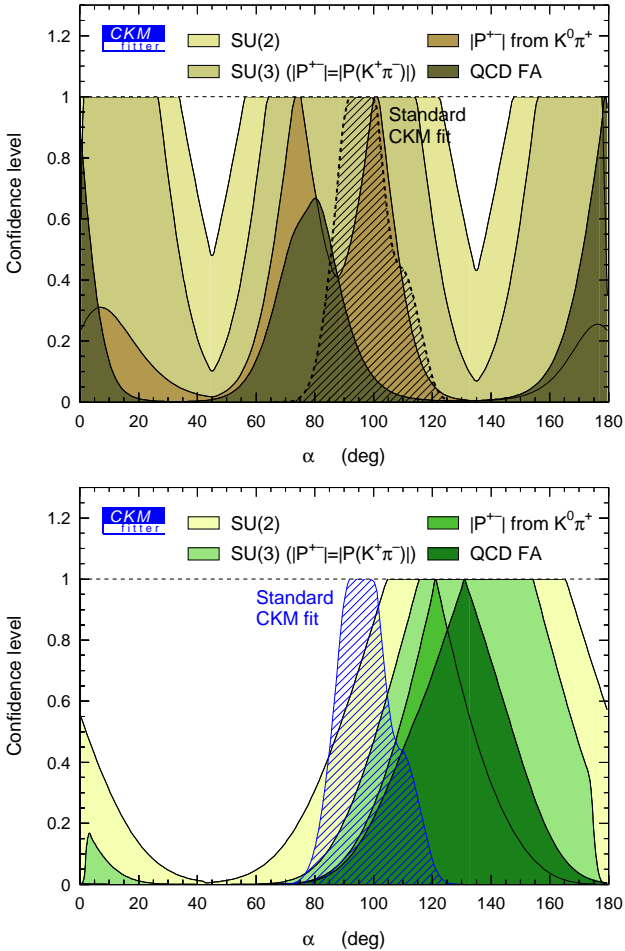} 
}
  \caption[.]{\label{fig:alpha}\em
        Confidence levels in $\alpha$ through various theoretical 
        frameworks and using $\Cpipi$ and $\Spipi$ \babar\ (top) 
        and Belle (bottom) measurements
        and world average branching
        fractions of $B\rar hh^\prime$ decays. The radius
	$(\rhobar^2+\etabar^2)^{1/2}$ has been confined using in addition
	$|V_{ub}|$. Overlaid
        (hatched area) is the prediction from the 
        standard CKM fit.}
\end{figure}

The constraints on $\alpha$, obtained in the frameworks discussed 
above, from the $\Spipi$ and $\Cpipi$ measurements of \babar\ and Belle
are shown in Fig.\ref{fig:alpha}. They are compared with the constraint 
obtained with the standard CKM fit \cite{CKMfitter}.

Due to the large upper limit for $\BRpizpiz$, potentially 
caused by the presence of a signal, there are essentially 
no constraints from the SU(2) analysis. Using in addition SU(3) 
one begins to weakly rule out regions of $\alpha$ and 
even more, non-trivial information can be obtained when estimating 
\boldmath$|\Ppipi|$ from $B^{+}\rar K^0\pi^+$. Nevertheless,
no theoretical errors have been assigned here with 
respect to the dynamical assumptions made. 
Only when using the highly predictive QCD factorisation 
for $|\Ppipi/\Tpipi|$ and $\deltapipi$ relatively stringent
constraints on $\alpha$  are obtained,
which with present statistics are already competitive with 
those obtained from the standard CKM fit and are found to be in
reasonnable agreement.

\begin{figure}[h]
\hbox to\hsize{\hss
\includegraphics[width=\hsize]{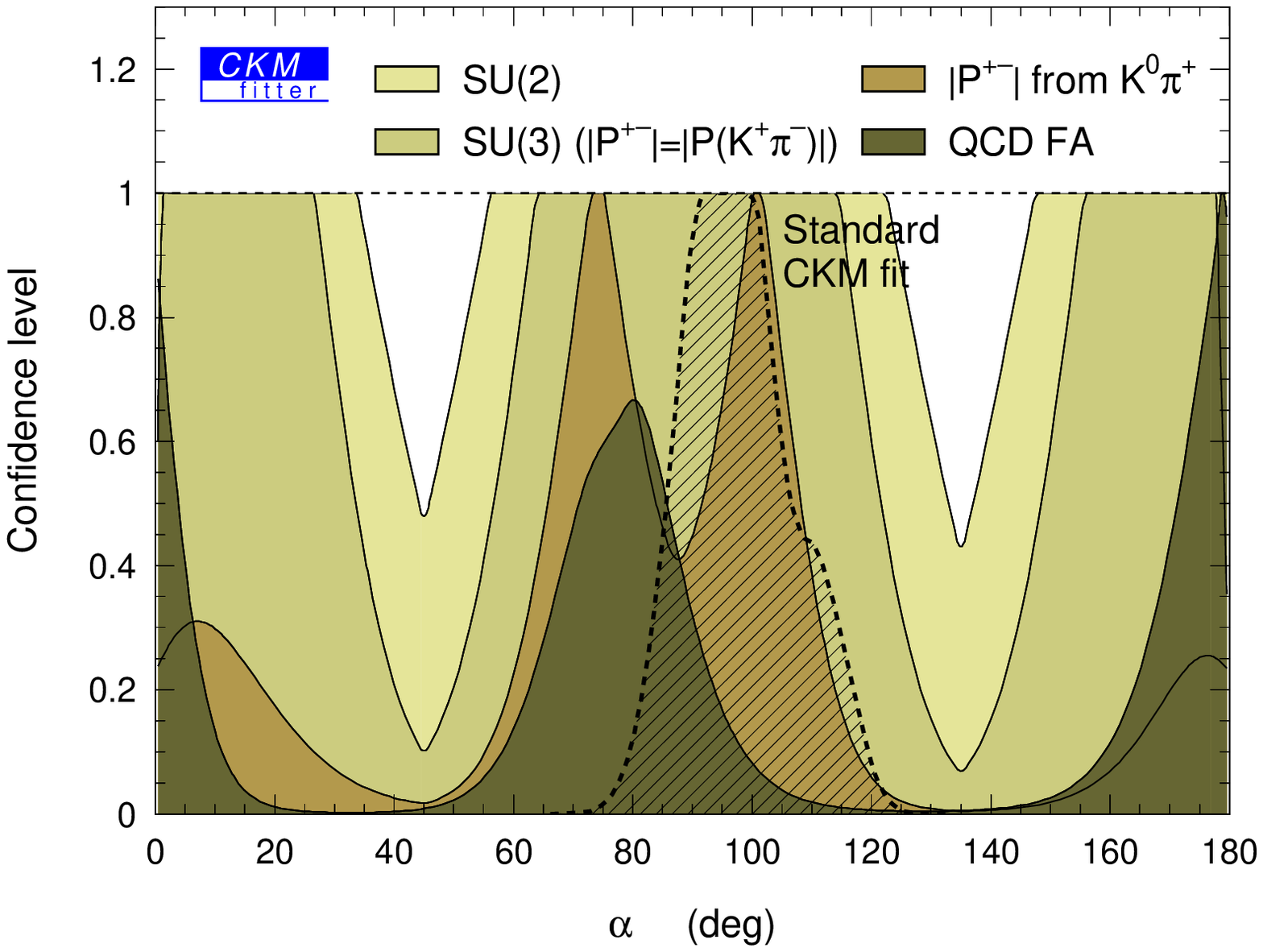} 
}
\end{figure}
\vspace{-1.5cm}

\begin{figure}
\hbox to\hsize{\hss
\includegraphics[width=\hsize]{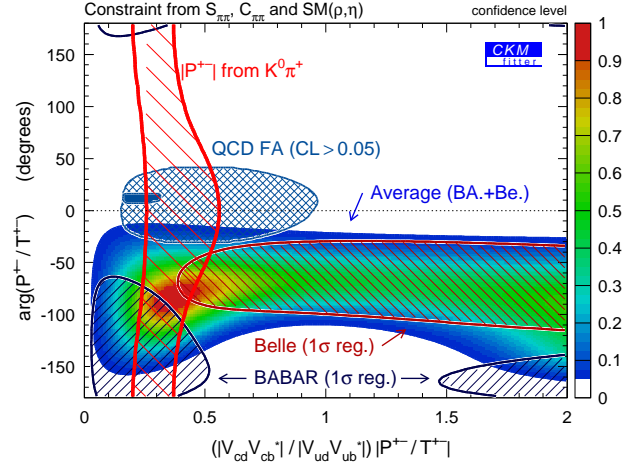}
}
  \vspace{0.0cm}
  \caption[.]{\label{fig:pot}\em
        Constraints on the penguin-to-tree ratio
        $|\Ppipi/\Tpipi|$ and the relative strong phase
        $\deltapipi$. The shaded
        (coloured) region gives the confidence levels obtained from 
        a fit using the isospin relation between amplitude and the
 	$\Cpipi$, $\Spipi$ measurements of \babar\ and Belle (averaged).
        Also shown are the 
        $5\%$~CL contours  corresponding to the approaches presented in
	subsection \ref{sec:naiveFA} and \ref{sec:QCDFA}. For QCD FA, the computation 
	results are shown with the non-factorizable contribution
	parametrization fixed (small elliptical area) and free (large
	elliptical area).}
\end{figure}

One can invert the point-of-view and constrain the QCD unknowns 
penguin-to-tree ratio $|\Ppipi/\Tpipi|$  and its phase 
instead of the CKM parameters.
This assumes that the measurements of $\Spipi$ and $\Cpipi$ 
are in agreement with the constraints obtained on $\rhobar$ and 
$\etabar$ in the standard CKM fit, \ie, no new physics comes into 
play. The resulting confidence levels are shown in Fig.~\ref{fig:pot}. 
Since the standard CKM fit constraints $\etabar$ to be positive 
(and thus $\gamma > 0$), the preferred values for $\deltapipi$ 
are negative to accomodate for negative values of $\Cpipi$. On the
other hand, QCD FA predicts a small positive $\deltapipi$ value.
Unfortunately, when the parametrization of the non-factorizable 
contributions is left free, QCD FA loses its predictible 
power\footnote{We remind that the experimental uncertainty in the
present knowledge of $\rhobar$ and $\etabar$ also propagates into
the $\Ppipi/\Tpipi$ calculation with QCD FA and that the theoretical
errors are added linearly.}
(hatched elliptical area in Fig.~\ref{fig:pot}).


\section{Prospects}

  \begin{figure}
\hbox to\hsize{\hss
\includegraphics[width=\hsize] {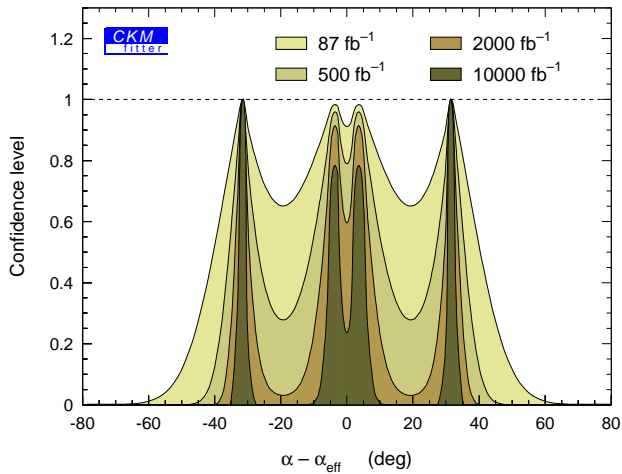}}
  \vspace{0.0cm}
  \caption[.]{\label{fig:dalpha}\em
        The residual $\dalpha$ for the ICHEP02 
        central values (see Table~\ref{tab:BRcompilation})
        and errors as well as hypothetical luminosities of 
        \babar, Belle and future B-factories (always keeping the 
        central values unchanged). Due to finite 
        binning only, the central peaks do not reach until 
        $CL=1$. The full isospin analysis is employed.}
\end{figure}

The ultimate goal of the experimental effort should be the 
unequivocal determination of $\alpha$ 
without relying on model dependent theoretical descriptions but
rather performing the full isospin analysis. In the following study,
present central values of branching fractions given in Table 
\ref{tab:BRcompilation} are assumed and the \babar\ values of $\Cpipi$ 
and $\Spipi$ are used.
\vs
Figure~\ref{fig:dalpha} shows the residual $\dalpha$ in four cases: 
$(i)$ the current statistics (but
assuming the individual rates $B^0({\overline B}^0)\rar\pi^0\pi^0$
have been measured, see below), $(ii)$ a luminosity 
of $500~{\rm fb}^{-1}$ reached by \babar\ in a few years from now,
$(iii)$ hypothetical $2000~{\rm fb}^{-1}$ reached by the first
generation $B$-factories at the end of their running period and, $(iv)$, 
$10000~{\rm fb}^{-1}$ collected by a second generation $B$ factory
after a few years of running. We assume that direct CPV is absent 
($|\lambda_{\pi^0\pi^0}|=1$) leading to the symmetric solutions 
shown in Fig.~\ref{fig:dalpha}. Mistag rates and the dilution from 
the time-integrated measurement have been taken into account in the 
extrapolation. For missing $B^0\rar\pi^0\pi^0$ flavour information 
(as currently the case), only the outer borders of the curves can 
be obtained from the isospin analysis, while the inner structure 
remains unresolved because the relative strong phase
is unconstrained. The inner structure reveals the remnants of the 
four-fold ambiguity of the full isospin analysis in the 
range $\alpha\in\{0,2\pi\}$. For this setup of central values, only 
luminosities of the order of $10~{\rm ab}^{-1}$ allow separation 
of the solutions. Without a separation, the overall allowed 
region exceeds the uncertainty of the SM fit.

\section{Conclusion}

We have studied various strategies proposed in the literature 
to interpret the time-dependent asymmetry measured in 
$B^0\rar\pi^+\pi^-$ decays in terms of CKM parameters.
\vs
At present, significant constraints on the penguin pollution are 
only provided by theoretical calculations predicting the complex
value of $\Ppipi/\Tpipi$ such as QCD FA. However, a 
confrontation of these calculations with experimental data of better 
statistics is still to come. 
\vs
The strategy proposed by Gronau and Rosner using $B^{+}\rar K^0\pi^+$
within some dynamical assumptions provides a qualitative interpretation.
The SU(2) bounds do not lead to useful constraints on 
$|\dalpha|$, since the upper limit 
on $\BRpizpiz$ is rather large. Not very well-known is the fact that 
a much better penguin bound is obtained from a strategy proposed by 
Charles, who uses the branching fraction of the penguin-dominated 
$B^0\rar K^+\pi^-$ and assumes SU(3) as well as weak additional 
theoretical presumptions. Unfortunately, at present, also in this case the 
constraint on $|\dalpha|$  is 
not stringent enough to measure $\alpha$. 
\vs
The present experimental information on $\BRpizpiz$ is compatible
with a branching ratio of the order $(1-2)\cdot10^{-6}$. In this case, 
it would be possible to perform the full isospin analysis as proposed by 
Gronau and London and to extract the penguin contribution from data. 
However, to constrain $|\dalpha|$ with good precision and to separate 
the ambiguities would require an integrated luminosity 
of $\sim 10~{\rm ab}^{-1}$.


\end{document}